\documentclass[twocolumn,showpacs,preprintnumbers,amsmath,amssymb]{revtex4}

\usepackage{graphicx}
\usepackage{epsfig}
\usepackage{dcolumn}
\usepackage{bm}

\begin{document}

\title{Probing the Nuclear Neutron Skin by Low-Energy Dipole Modes }

\author{N. Tsoneva}
 \altaffiliation[Also at ]{Institute for Nuclear Research and Nuclear Energy, 1784 Sofia, Bulgaria}
 \email{Nadia.Tsoneva@theo.physik.uni-giessen.de}
\author{H. Lenske}
\affiliation{Institut f\"ur Theoretische Physik, Universit\"at
Gie\ss en, Heinrich-Buff-Ring 16, D-35392 Gie\ss en, Germany}

\author{Ch. Stoyanov}
\affiliation{Institute for Nuclear Research and Nuclear Energy, 1784 Sofia, Bulgaria }

\date{\today}

\begin{abstract}
Dipole excitations  below the neutron threshold in neutron rich Sn
isotopes are studied theoretically in the Quasiparticle-Phonon
Model with Hartree-Fock-Bogoliubov single particle input. 
Of special interest are
the low-lying two-phonon $1^{-}$ states and the Pygmy Dipole
Resonance (PDR). The evolution of low-energy dipole excitations
with neutron excess is investigated over the Sn isotopic chain
including the experimentally unknown regions close to $^{132}$Sn.
A dependence of the PDR strengths and centroid energies on
the neutron skin thickness is found. Despite significant
multi-phonon contributions to mean energies and transition
strengths, the PDR states retain their one-phonon character. The
fragmentation pattern is reduced with increasing neutron excess
towards the N=82 shell closure which will be of advantage for
future experimental work.
\end{abstract}

\maketitle

A genuine feature of neutron-rich nuclei is the appearance of
low-energy electric dipole strength, seen recently in
high-precision photon scattering experiments already in stable
nuclei with small \cite{Ca-pdr,Ba-pdr} and moderate \cite{Pb-pdr}
neutron excess.  These so-called {\it Pygmy Dipole Resonances
(PDR)} are observed as a clustering of states close to the neutron
threshold which in presently accessible stable medium- and
heavy-mass nuclei is at excitation energies E$_x\sim5.5-$8~MeV.
In a first attempt for experimental investigations on the dipole strength distribution in exotic nuclei with radioactive beams recently performed in MSU and GSI the neutron-rich (stable and unstable) oxygen isotopes are studied. A dipole strength is observed below \cite{Try} and above \cite{Lei} the neutron threshold. The measurements of the latest in the energy region 5$-$8 MeV for $^{18,20}$O 'are consistent with that for other nuclei that may exhibit the Pygmy Dipole Resonance'.
Although carrying only a small fraction of the full dipole
strength the PDR states are of particular interest because they are
expected to reflect the motion of the neutron skin against the
core of normal nuclear matter.  In order to obtain more direct
evidence for such a peculiar mode studies of low-energy dipole
excitations over sufficiently long isotopic chains are necessary.
Here, we present results of an exploratory theoretical
investigation for the neutron-rich Sn isotopes in experimentally
less or even unknown regions at neutron numbers N=$70-82$. 
As discussed in \cite{Pb-pdr}, the nature of the low-energy dipole
strength differs significantly from the isovector Giant Dipole
Resonance (GDR) mode where proton and neutron fluids as a whole
move against each other. Moreover, the PDR mode has to be
distinguished from the other known low-energy isoscalar dipole
excitation, namely the two-phonon $1^-$ states resulting from the
anharmonic interactions of the lowest $2^+$ and $3^-$ states in a
nucleus. The anharmonicities are reflecting the intrinsic
fermionic structure of the nuclear phonons thus deviating from
ideal bosons. These competing effects are taken into account in an
appropriate way by the Quasiparticle Phonon Model (QPM)\cite{Sol}.
Applications of the QPM to low-energy dipole strength
\cite{Pon2,Gri1,Pon3,Pon1} and also in the recent PDR
investigations in $^{208}$Pb \cite{Pb-pdr} have led to very good
descriptions of data thus giving confidence on the reliability of
the model for such investigations. 

The QPM approach is at present the only method allowing for
sufficiently large configuration space such that a unified
description of low-energy single and multiple phonon states and
the GDR is feasible. In the one-phonon sector 
the Quasiparticle Random-Phase Approximation (QRPA) is recovered as a
well understood limiting case if the quasi-boson approximation is
used. Such a unified treatment is exactly what is required in
order to separate the multi-phonon and the genuine PDR $1^{-}$
strengths in a meaningful way. For the aim of this paper we use
the standard form of QPM \cite{Sol}, i.e. approximating residual
interactions in terms of separable multipole-multipole
interactions with empirical coupling constants, see e.g.
\cite{Vdo,Vor}. The isoscalar quadrupole and octupole
coupling constants are chosen to reproduce the
experimental energies and electromagnetic transitions of the
2$^{+}_1$ and 3$^{-}_1$ states. 
Since for A$<$126 where the data are available the coupling constants vary smoothly with neutron number on a level of less than 5$\%$ extrapolations into the unexplored mass region towards $^{132}$Sn can be done safely. 

Since single particle energies and a reliable description of
ground state properties in general are critical quantities for
extrapolations of QRPA and QPM calculations into unknown mass
regions here we put special emphasis on the mean-field part.
Because of the numerical constraints set by the QPM a
semi-microscopic approach is chosen. Taking advantage of the
Kohn-Sham theorem \cite{KS} of Density Functional Theory (DFT) the total
binding energy $B(A)$ is expressed as an integral over an energy
density functional with (quantal) kinetic ($\tau$) and self-energy
parts, respectively,
\begin{equation}
B(A)=\int{d^3r\left( \tau(\rho)+\frac{1}{2}\rho
U(\rho)\right)}+E_{pair}
\end{equation}
\[
=\sum_j{v_j^2\left(e_j-<\Sigma>_j+\frac{1}{2}<U>_j\right)}+E_{pair}
\quad ,
\]
and pairing contributions are indicated by $E_{pair}$. The second
relation is obtained from Hartree-Fock-Bogoliubov (HFB) theory with 
occupancies $v^2_j$ and
potential energies $<U>_j$ of the occupied levels $j$, see e.g.
\cite{Len,Hofmann}. Above, $U(\rho)$ is the proper self-energy,
i.e. not including the rearrangement contributions from the
intrinsic density dependence of nuclear interactions
\cite{Hofmann,Hofmann01}. Hence, $U(\rho)$ has to be distinguished
from the effective self-energy obtained by variation
\begin{equation}
\Sigma(\rho)=\frac{1}{2}\frac{\partial \rho U(\rho)}{\partial
\rho}
\end{equation}
and appearing in the single particle Schroedinger equation. In
order to keep the QPM calculations feasible we choose $\Sigma
\equiv U_{WS}$ to be of Wood-Saxon (WS) shape with adjustable
parameters. By inversion and observing that the densities and
potentials in a finite nucleus are naturally given parametrically
as functions of the radius $r$, we find
\begin{equation}
\rho(r) U(r)= -2 \int_r^\infty{ds \frac{\partial \rho(s)}{\partial
s} U_{WS}(s)} \quad .
\end{equation}
Evaluating these relations with the microscopic proton and neutron
densities obtained by solving the Schroedinger equation with
$U_{WS}$ the potential $U(\rho)$ is the self-consistently derived
reduced self-energy entering e.g. into the binding energy.

\begin{figure}
\epsfig{file=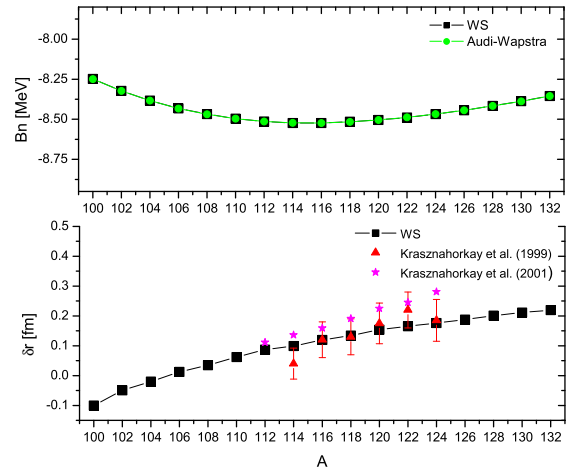,width=\linewidth}
\caption{\label{fig:fig1} Ground state properties of the Sn
isotopes. In the upper panel the total nuclear binding energies
per particle calculated with the adjusted WS potentials are compared to the values from the Audi-Wapstra
compilation \protect\cite{Audi95}. In the lower panel the calculated with the adjusted WS potentials
differences of proton and neutron rms radii are compared to the
experimental values obtained by Krasznahorkay et al. from charge exchange reactions in
ref. \protect\cite{Sn-skin1,Sn-skin2}.}\label{groundstates}
\end{figure}

In practice, for a given nucleus of mass $A$ the depth of the
central and spin-orbit potentials, radius and diffusivity
parameters of $U_{WS}$ are adjusted separately for protons and
neutrons to the corresponding single particle separation energies,
the total binding energy \cite{Audi95}, the charge radii and
(relative) differences of proton and neutron root-mean-square
(RMS) radii,
\begin{equation}\label{dr}
\delta r=\sqrt{<r^2>_n}-\sqrt{<r^2>_p} \quad ,
\end{equation}
from our previous HFB calculations \cite{Hofmann,Hofmann01}.
The theoretically obtained RMS radii are compared to those determined from charge exchange reactions by Krasznahorkay et al. \cite{Sn-skin1,Sn-skin2} for a number of Sn isotopes. 
However, since the measurements did not provide absolute $\delta r $ values the 
data were normalized in \cite{Sn-skin1,Sn-skin2} to theoretical predictions. 
Hence, strictly spoken only the relative mass dependence of the experimental
results shown in Fig.\ref{fig:fig1} is of significance.
The HFB energies were scaled by an average effective mass of $m^*/m=0.68$ 
thus removing
the known problem of unrealistically large HFB level spacings at
the Fermi surface. The approach sketched above leads to very
satisfactory results on binding energies and proton-neutron
RMS-differences as shown in Fig.\ref{fig:fig1} for the Sn
isotopes. A smooth dependence of the parameters on $A$ is found
which supports the reliability of the method.

In the QPM calculations the structure of the excited states is described
by a wave function including up to three-phonon configurations \cite{Gri1}
which are built from a basis of QRPA states of natural parity
excitations with J$^\pi=1^{-}\div 5^-$. Since the
one-phonon configurations up to E$_x$=20~MeV are considered the 
core polarization contributions to the transitions of
the low-lying 1$^{-}$ states are taken into account explicitly.
Hence, we do not need to introduce effective charges. The two- and
three-phonon configurations are truncated at E$_x$=4.5~MeV for the
calculation of the quadrupole-octupole 1$^{-}$ state for the
proper comparison with the available Nuclear Resonance Fluorescence (NRF)
 data \cite{Pon2}. For the
QPM calculations between 4.5$\div$8~MeV the two- and three-phonon
basis is limited to states up to E$_x$=8.5~MeV and 8~MeV,
respectively.

As an important prerequisite for the reliability of
the model wave functions we consider first the lowest J$^{\pi}$=2$^+$ and J$^{\pi}$=3$^-$
states in the $^{120\div130}$Sn isotopes. The QPM energies and transition probabilities of these states are presented in Table I and
compared to the known experimental data. The well reproduced
properties of the 2$^{+}_{1}$ and 
3$^{-}_{1}$ states are important for a trustable
description of the two-phonon quadrupole-octupole J$^{\pi}$=1$^{-}_{1}$ state
which is of our particular interest. The experimental data on excitation
energies and transition strengths \cite{NDT}, available for
$^{120\div130}$Sn, are well described by the QPM as seen from Table I.

\begin{table}

\caption{\label{tab1}QPM results for the energies and the reduced
B(E1), B(E2) and B(E3) transition probabilities of the first
1$^{-}$, 2$^{+}$ and 3$^{-}$ states in $^{120\div130}$Sn isotopes.
A comparison with the experimental data \cite{Pon2} is presented.}
\begin{tabular}{cccccclcc}
\hline\hline Nucl.&  & \multicolumn{2}{l}{Energy} & Trans. & 
\multicolumn{3}{l}{B(E1; I$_\nu ^\pi \rightarrow$ J$_{\nu'} ^{\pi'}$) [10$^{-3}    $ e$^2$fm$^2$]} \\
&  & [MeV] &  &  &\multicolumn{3}{l}{B(E2; I$_\nu ^\pi \rightarrow$ J$_{\nu'} ^{\pi'}$) [10$^4 $ e$^2$fm$^4$]} \\
&  &  &  &  & \multicolumn{3}{l}{B(E3; I$_\nu ^\pi \rightarrow$ J$_{\nu'} ^{\pi'}$) [10$^6 $ e$^2$fm$^6$]} \\
\hline &J$_{\nu'} ^{\pi'}$ & Exp. & QPM & E$\lambda$ & I$_\nu ^\pi
$ &  Exp. & QPM \\ \cline{2-8}
&  &  &  &  \\
$^{120}$Sn & 2$_1^{+}$ & 1.171 & 1.171 & E2 & 0$^{+}_1$& 0.200(3)& 0.193\\
&  &  &  & E1 & 3$_1^{-}$& 2.02(17) & 1.82\\
& 3$_1^{-}$ & 2.401 & 2.424 & E3 & 0$^{+}_1$& 0.115(15)
& 0.110\\
& 1$_1^{-}$ & 3.279 & 3.203 & E1 & 0$^{+}_1$&7.60(51)  & 7.6 \\
$^{122}$Sn & 2$_1^{+}$ & 1.141 & 1.137 & E2 & 0$^{+}_1$& 0.194(11) & 0.190\\
&  &  &  & E1 & 3$_1^{-}$&2.24(14)  & 2.06\\
& 3$_1^{-}$ & 2.493 & 2.486 & E3 & 0$^{+}_1$& 0.092(10)
& 0.099 \\
& 1$_1^{-}$ & 3.359 & 3.281 & E1 & 0$^{+}_1$&7.16(54)  &7.02  \\
$^{124}$Sn & 2$_1^{+}$ & 1.132 & 1.133 & E2 & 0$^{+}_1$& 0.166(4) & 0.174  \\
&  &  &  & E1 & 3$_1^{-}$& 2.02(16) & 1.98\\
& 3$_1^{-}$ & 2.614 & 2.645 & E3 & 0$^{+}_1$& 0.073(10)
& 0.087  \\
& 1$_1^{-}$ & 3.490 & 3.549 & E1 & 0$^{+}_1$& 6.08(66) & 6.27 \\
$^{126}$Sn & 2$_1^{+}$ & 1.141 & 1.151 & E2 & 0$^{+}_1$& -& 0.140  \\
&  &  &  & E1 & 3$_1^{-}$& - & 1.74\\
& 3$_1^{-}$ & 2.720 & 2.792 & E3 & 0$^{+}_1$& - & 0.079 \\
& 1$_1^{-}$ & - & 3.856 & E1 & 0$^{+}_1$& - & 5.8\\
$^{128}$Sn & 2$_1^{+}$ & 1.168 & 1.154 & E2 & 0$^{+}_1$& - & 0.097 \\
&  &  &  & E1 & 3$_1^{-}$& - & 1.07\\
& 3$_1^{-}$ & - & 2.849 & E3 & 0$^{+}_1$& - & 0.081 \\
& 1$_1^{-}$ & - & 4.115 & E1 & 0$^{+}_1$& - & 5.56\\
$^{130}$Sn & 2$_1^{+}$ & 1.221 & 1.204 & E2 & 0$^{+}_1$ & - & 0.066 \\
&  &  &  & E1 & 3$_1^{-}$& - & 1.11\\
& 3$_1^{-}$ & - & 2.861 & E3 & 0$^{+}_1$ & - & 0.098\\
& 1$_1^{-}$ & - & 4.094 & E1 & 0$^{+}_1$ & - & 5.53 \\
\hline\hline
\end{tabular}
\end{table}

The QPM predicts for the lowest-lying 1$^{-}$ states excitation energies
varying from E$_x$=3.203~MeV to E$_x$=4.115~MeV in
$^{120\div130}$Sn (see Table I). In these isotopes  
the 1$^{-}_{1}$ states are at a level of about 90$\%$ predominantly given 
by two-phonon quadrupole-octupole configurations. Since with increasing
neutron number the energy of the latter becomes larger, certain
other, higher-lying, two-phonon configurations also start to
contribute. The three-phonon configurations are most important for
the lower-mass Sn isotopes because of their open shell structure.
The reduction correlates with the decrease of collectivity when
approaching the N=82 shell closure. 
The QPM results for the {\it boson forbidden} E1 transitions
\cite{Pon3} from the two-phonon 1$^{-}$ state to the ground state
and between the 3$^{-}_{1}$ and 2$^{+}_{1}$ states are also shown
in Table \ref{tab1}. The experimental data available for
$^{116}$Sn$\div$$^{124}$Sn \cite{Pon2} are well described.

\begin{table*}
\caption{\label{tab2}Reduced transition probabilities and mean
energies of the PDR modes in $^{120}$Sn$\div$$^{132}$Sn calculated
within QRPA are presented. The contribution of the PDR mode in the
Energy Weighted Sum Rule (EWSR) is given in \%. For $^{124}$Sn the
calculations are compared with the experimental data}.
\begin{ruledtabular}
\begin{tabular}{llllllll}

Nucleus & \multicolumn{3}{l}{$\left\langle E \right\rangle$ [MeV]} &
\multicolumn{3}{l}{$\sum B(E1)\uparrow  [e^2fm^2]$} & EWSR (\%) \\ \hline
& Exp. & QRPA & QPM & Exp. & QRPA &QPM  \\ \cline{2-4} \cline{4-7}
$^{120}$Sn & - & 6.839& 6.920  & - & 0.142 & 0.289  &.185 \\
$^{122}$Sn & - & 6.695& 6.708  & - & 0.181 & 0.347 &.228 \\
$^{124}$Sn & 6.5$^{[10]}$& 6.544 &6.641  &0.345(43)$^{[10]}$ & 0.219 &0.398 &.267 \\
$^{126}$Sn & - & 6.330 &6.442  & - & 0.261 & 0.416   &.304 \\
$^{128}$Sn & - & 6.100&6.210  & - & 0.281 &0.434  &.313 \\
$^{130}$Sn & - & 5.940 &6.150  & - & 0.337 &0.495  &.361 \\
$^{132}$Sn & - & 5.900 & - & - & 0.360 & -  &.380 \\
\end{tabular}
\end{ruledtabular}
\end{table*}

Of central interest for this work are the 1$^{-}$ states above the
two-phonon dipole state and below the neutron threshold and their
evolution with the neutron excess. These states are eventually to
be identified as PDR modes related to excitations of the neutron
skin as found in $^{208}$Pb \cite{Pb-pdr}. In fact, in all nuclei
considered here the first three 1$^{-}$ QRPA states contribute to
this energy region. They are rather well separated 
by a energy gap of more than 1.3~MeV from other
higher-lying one-phonon 1$^{-}$ states which are more likely to 
belong to the low-energy tail of the GDR. 
QRPA and QPM results on the total PDR strengths
and average centroid energies (defined by the ratios of energy and
non-energy weighted sum rules) in Sn isotopes are compared in
Tab.\ref{tab2}. As a general result we find a correlation of the
total PDR strength and the neutron excess as reflected by the
increase of the total PDR strength with mass number. The PDR centroid
energies, however, decrease with mass number. They are only
weakly affected by multi-phonon admixtures.

The calculated energies and transition strengths are in a good
agreement with experimental data \cite{Pon1}. The detailed
analysis of the E1 strength distribution over the 1$^{-}$ excited
states in $^{120\div130}$Sn isotopes reveals that the major part
of the strength is concentrated in the states with relatively large
one-phonon $[1^{-}_{2}]$ or $[1^{-}_{3}]$ components. For the
total PDR strengths the differences between the QRPA and QPM are
significant (see Table \ref{tab2}). Although the contribution of
the higher-lying one-phonon components to the wave function structure 
of the low-lying 1$^-$ states is small their influence on the total E1 
transition is important because of partial admixtures of the large
collective one-phonon matrix elements from the GDR region. The QPM B(E1)
values are enhanced significantly by factors ranging from about 2
to 1.5 for $A=120$ to $A=130$, respectively. Thus, the PDR
transition strengths contain a considerable fraction of strength
from higher-lying (GDR) states, virtually admixed through the
anharmonic QPM interactions. These admixtures, in fact,
overcompensate the loss of ground state transition strength
resulting from the dissipation of the one-phonon states into the
multi-phonon components. In $^{122}$Sn and $^{130}$Sn the QPM
calculations exhaust about 85$\%$ of the PDR one-phonon strength
in the bound energy region. In Table \ref{tab2} also the relative
contribution of the PDR to the Energy Weighted Sum Rule (EWSR) is
given.

An interesting question is to relate the PDR modes to the neutron
skin. While in \cite{Pb-pdr} this was done on the level of
transition densities here we analyze the PDR strength
distributions as functions of $\delta r$, eq. \ref{dr}. We find
two well separated regions with a sudden increase in the slope
beyond $A=126$ where $\delta r$ starts to exceed 0.2~fm. Hence, a
clear signal on the thickness of the neutron skin is obtained
allowing to extract information on nuclear shapes from PDR data.

\begin{figure}
\includegraphics{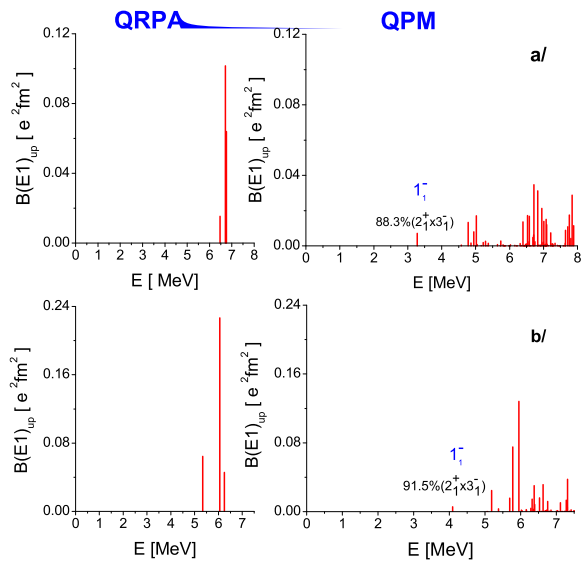}
\caption{\label{fig:fig2}QRPA and QPM calculations of the E1
strength distribution in a/$^{122}$Sn and b/$^{130}$Sn below the
neutron threshold.}
\end{figure}

The QRPA and QPM dipole strength distributions below neutron
threshold in $^{122}$Sn and $^{130}$Sn are compared in
Fig.\ref{fig:fig2}. The lowest QPM 1$^{-}$ states do not have QRPA
counterparts because they are the quadrupole-octupole two-phonon
states. The QPM anharmonicities introduce a considerable
fragmentation -- in these cases up to about 80 states. Compared to
\cite{Pon1} we find less fragmentation because our two- and
three-phonon configuration spaces are somewhat smaller. However, 
the obtained  for PDR total strength and centroid energy is in
very reasonable agreement with the available experimental data in
$^{124}$Sn (see Table \ref{tab2}).

The lowest QRPA 1$^{-}$ states are mainly given by pure neutron 
two-quasiparticle excitations 
with small admixtures of higher-lying two-quasiparticle neutron and proton configurations 
on a level of a few percent or less.
The dominance of neutron excitations in this energy region
is in agreement with the results of
previous PDR studies in Sn and other nuclei by relativistic random-phase approximation (RRPA) \cite{Paar}, DFT \cite{CA} and QPM \cite{Pb-pdr}. The remarkable stability of the wave function structure over the isotopic chain resembles
what is found theoretically also for the GDR. However, we
emphasize that the PDR states are of a much stronger isoscalar (or
eventually mixed isospin) content than the GDR, see also
\cite{Pb-pdr}. Hence, the PDR states cannot be considered as
simply being the low-energy tail of the {\it isovector} GDR,
despite the anharmonic admixtures discussed above. Rather, these
dipole modes are of a genuine character which cannot be deduced by
extrapolations from the GDR region. This conclusion is supported
by the differences in transition densities and velocity fields
discussed in \cite{Pb-pdr}.

In summary, neutron rich tin isotopes were studied in a
semi-phenomenological approach combining HFB and QPM theory. The
results confirm the success of such a microscopically inspired
description. An important step in understanding the dipole spectra
is to disentangle the PDR states from the low-energy two-phonon
dipole states, achieved here by using the QPM approach with up to
three-phonon configurations. For $^{120}$Sn $\div$ $^{132}$Sn we
obtained low-energy dipole strength concentrated in a narrow
energy interval such that a {\it pygmy dipole resonance} (PDR) can
be identified. The evolution of the PDR strength distribution with
neutron excess shows that transition strengths and energy
locations are indeed closely correlated with the neutron skin.
Hence, measurements of PDR strength distributions will provide
information on nuclear shapes far off stability.

\noindent This work is supported by DFG, contract Le439/5 and Bulgarian Foundation for Scientific Research, contract Ph 1311.


\end{document}